\documentclass[preprint,11pt]{elsarticle}

 \usepackage{epsfig}

\usepackage{amssymb}
\usepackage{graphics,amssymb,amsmath}
\usepackage{graphicx}
\usepackage{subfigure}
\newproof{pf}{Proof}
\newdefinition{rmk}{Remark}
\newtheorem{thm}{Theorem}
\newdefinition{prop}{Proposition}
\begin{document}

\begin{frontmatter}



\title{Exponentiated Weibull-Geometric Distribution and its Applications}


\author{Eisa Mahmoudi\corref{cor1}}
\ead{emahmoudi@yazduni.ac.ir}
\author{Mitra Shiran}
%\ead[url]{http://www.elsevier.com}

\cortext[cor1]{Corresponding author}

\address{Department of Statistics, Yazd University,
P.O. Box 89175-741, Yazd, Iran}
%\address[rvt]{Department of Statistics, Yazd University,
 %P.O. Box 89175-741, Yazd, Iran}

\begin{abstract}
In this paper a new lifetime distribution, which is called the
exponentiated Weibull-geometric (EWG) distribution, is introduced.
This new distribution obtained by compounding the exponentiated
Weibull and geometric distributions. The EWG distribution includes
as special cases the generalized exponential-geometric (GEG),
complementary Weibull-geometric (CWG), complementary
exponential-geometric (CEG), exponentiated Rayleigh-geometric (ERG)
and Rayleigh-geometric (RG) distributions.

The hazard function of the EWG distribution can be decreasing,
increasing, bathtub-shaped and unimodal among others. Several
properties of the EWG distribution such as quantiles and moments,
maximum likelihood estimation procedure via an EM-algorithm,
R\'{e}nyi and Shannon entropies, moments of order statistics,
residual life function and probability weighted moments are studied
in this paper. In the end, we give two applications with real data
sets to show the flexibility of the new distribution.

\end{abstract}


\end{frontmatter}


\section{Introduction}

The Weibull and exponentiated Weibull (EW) distributions in spite of
their simplicity in solving many problems in lifetime and
reliability studies, do not provide a reasonable parametric fit to
some practical applications.

Recently, attempts have been made to define new families of
probability distributions that extend well-known families of
distributions and at the same time provide great flexibility in
modeling data in practice. One such class of distributions generated
by compounding the well-known lifetime distributions such as
exponential, Weibull, generalized exponential, exponentiated Weibull
and etc with some discrete distributions such as binomial,
geometric, zero-truncated Poisson, logarithmic and the power series
distributions in general. The non-negative random variable $Y$
denoting the lifetime of such a system is defined by ${{Y=\min
}_{1\le i\le N} X_i\ }$ or ${{Y=\max}_{1\le i\le N} X_i\ }$, where
the distribution of $X_i$ belongs to one of the lifetime
distributions and the random variable $N$ can have some discrete
distributions, mentioned above.

This new class of distributions has been received considerable
attention over the last years. The exponential-geometric (EG),
exponential-Poisson (EP), exponential- logarithmic (EL), exponential
power series (EPS), Weibull-geometric (WG), Weibull-power series
(WPS), exponentiated exponential Poisson (EEP), complementary
exponential-geometric (CEG), two-parameter Poisson-exponential,
generalized exponential power series (GEPS), exponentiated
Weibull-Poisson (EWP) and generalized inverse Weibull-Poisson (GIWP)
distributions were introduced and studied by Adamidis and Loukas
\cite{Adamidis }, Kus \cite{Kus }, Tahmasbi and Rezaei
\cite{Tahmasbi }, Chahkandi and Ganjali \cite{Chahkandi},
Barreto-Souza et al. \cite{Barreto-Souza2011 }, Morais and
Barreto-Souza \cite{Morais }, Barreto-Souza and Cribari-Neto
\cite{Barreto-Souza2009 }, Louzada-Neto et al. \cite{Louzada},
Cancho et al. \cite{Cancho }, Mahmoudi and Jafari
\cite{Mahmoudi2011a }, Mahmoudi and Sepahdar \cite{Mahmoudi2011b }
and Mahmoudi and Torki \cite{Mahmoudi2011c }.

In this article, we propose a new four-parameter distribution,
referred to as the EWG distribution which contains as special
sub-models the generalized exponential-geometric (GEG),
complementary Weibull-geometric (CWG), complementary
exponential-geometric (CEG), exponentiated Rayleigh-geometric (ERG)
and Rayleigh-geometric (RG) distributions. The hazard function of
the EWG distribution can be decreasing, increasing, bathtub-shaped
and unimodal. Several properties of the EWG distribution such as
quantiles and moments, maximum likelihood estimation procedure via
an EM-algorithm, R\'{e}nyi and Shannon entropies, moments of order
statistics, residual life function and probability weighted moments
are studied in this paper.

The paper is organized as follows. In Section 2, we review the EW
distribution and its properties. In Section 3, we define the EWG
distribution. The density, survival and hazard rate functions and
some of their properties are given in this section. Section 4
provides a general expansion for the quantiles and moments of the
EWG distribution. Its moment generating function is derived in this
section. R\'{e}nyi and Shannon entropies of the EWG distribution are
given in Section 5. Section 6 provides the moments of order
statistics of the EWG distribution. Residual life function of the
EWG distribution is discussed in Section 7. In Section 8 we explain
probability weighted moments. Mean deviations from the mean and
median are derived in Section 9. Section 10 is devoted to the
Bonferroni and Lorenz curves of the EWG distribution. Estimation of
the parameters by maximum likelihood via an EM-algorithm and
inference for large sample are presented in Section 11. In Section
12, we studied some special sub-models of the EWG distribution.
Applications to real data sets are given in Section 13 and
conclusions are provided in Section 14.

\section{ Exponentiated Weibull distribution: A brief review }
Mudholkar and Srivastava \cite{Mudholkar1993 } introduced the EW
family as extension of the Weibull family, which contains
distributions with bathtub-shaped and unimodal failure rates besides
a broader class of monotone failure rates. One can see Mudholkar et
al. \cite{Mudholkar1995 }, Mudholkar and Huston \cite{Mudholkar1996
}, Gupta and Kundu \cite{Gupta2001 }, Nassar and Eissa \cite{Nassar}
and Choudhury \cite{Choudhury} for applications of the EW
distribution in reliability and survival studies.

The random variable $X$ has an EW distribution if its cumulative
distribution function (cdf) takes the form
\begin{equation}\label{cdf ExW}
F_{X}(x)=\left(1-e^{-(\beta x)^{\gamma}}\right)^{\alpha},~~x>0,
\end{equation}
where $\gamma>0$, $\alpha>0$ and $\beta>0$, which is denoted by
$EW(\alpha,\beta,\gamma)$. The corresponding probability density
function (pdf) is
\begin{equation}\label{pdf ExW}
f_X(x)=\alpha\gamma\beta^{\gamma}x^{\gamma-1}e^{-(\beta
x)^{\gamma}}\left(1-e^{-(\beta x)^{\gamma}}\right)^{\alpha-1}.
\end{equation}
The survival and hazard rate functions of the EW distribution are
\begin{equation*}
S(x)=1-\left(1-e^{-(\beta x)^{\gamma}}\right)^{\alpha},
\end{equation*}
and
\begin{equation*}
h(x)= \alpha\gamma\beta^{\gamma}x^{\gamma-1}e^{-(\beta
x)^{\gamma}}\left(1-e^{-(\beta
x)^{\gamma}}\right)^{\alpha-1}\left[1-\left(1-e^{-(\beta
x)^{\gamma}}\right)^{\alpha}\right]^{-1},
\end{equation*}
respectively. The $k$th moment about zero of the EW distribution is
given by
\begin{equation}\label{mean EW}
E(X^{k})=\alpha\beta^{-k}\Gamma\left(\frac{k}{\gamma}+1\right)\sum_{j=0}^{\infty}(-1)^j
{\alpha-1 \choose j}(j+1)^{-(\frac{k}{\gamma}+1)}.
\end{equation}
Note that for positive integer values of $\alpha$, the index $j$ in
previous sum stops at $\alpha-1$, and the above expression takes the
closed form
\begin{equation}
E(X^{k})=\alpha\beta^{-k}\Gamma\left(\frac{k}{\gamma}+1\right)A_{k}(\gamma),
\end{equation}
where
\begin{equation}\label{v.p}
A_{k}(\gamma)=1+\sum_{j=1}^{\alpha-1}(-1)^j {\alpha-1 \choose
j}(j+1)^{-(\frac{k}{\gamma}+1)},~~~k=1,2,3,\cdots,
\end{equation}
in which $\Gamma(.)$ denotes the gamma function (see, Nassar and
Eissa \cite{Nassar} for more details).

\section{ The EWG distribution }
Consider the random variable $X$ having the EW distribution where
its cdf and pdf are given in (\ref{cdf ExW}) and (\ref{pdf ExW}).\\
Given $N$, let $X_{1},\cdots,X_{N}$ be independent and identically
distributed (iid) random variables from EW distribution. Let the
random variable $N$ is distributed according to the geometric
distribution with pdf
\begin{equation*}
P(N=n)=(1-\theta)\theta^{n-1},~n=1,2 ,\cdots,~0\leq\theta<1.
\end{equation*}
Let $Y=\max(X_{1},\cdots,X_{N})$, then the conditional cdf of
$Y|N=n$ is given  by
\begin{equation}\label{dist y given N}
F_{Y|N=n}(y)=\left(1-e^{-(\beta y)^{\gamma}}\right)^{n\alpha},
\end{equation}
which is the EW distribution with parameters $n\alpha$, $\beta$,
$\gamma$, and denoted by EW$(n\alpha,\beta,\gamma)$. The
exponentiated Weibull-geometric (EWG) distribution, denoted by
EWG$(\alpha,\beta,\gamma,\theta)$, is defined by the marginal cdf of
$Y$, i.e.,
\begin{equation}\label{cdf EWG}
F_Y(y)=\frac{(1-\theta)\left(1-e^{-(\beta
y)^{\gamma}}\right)^{\alpha}}{1-\theta\left(1-e^{-(\beta
y)^{\gamma}}\right)^{\alpha}}.
\end{equation}
This new distribution includes some sub-models such as the
complementary exponential-geometric (CEG), generalized
exponential-geometric (GEG), complementary Weibull-geometric (CWG),
exponentiated Rayleigh-geometric (ERG) and Rayleigh-geometric (RG)
as special cases.
%\begin{rmk}
%Let $W=\min\{X_{i}\}^{N}_{i=1}$, then the cdf of $W$ is given by
%\begin{equation*}
%F_{W}(w)=,~~w>0.
%\end{equation*}
%If $\gamma=1$, then the cdf of $W=\min\{X_{i}\}^{N}_{i=1}$, in which
%$X_{i}$'s has the exponentiated Weibull distribution is obtained.
%\end{rmk}
The pdf of the EWG distribution is given by
\begin{equation}\label{pdf EWG}
f_Y(y)=\frac{(1-\theta)\alpha\gamma\beta^{\gamma}y^{\gamma-1}e^{-(\beta
y)^{\gamma}}\left(1-e^{-(\beta
y)^{\gamma}}\right)^{\alpha-1}}{\left[1-\theta\left(1-e^{-(\beta
y)^{\gamma}}\right)^{\alpha}\right]^2},
\end{equation}
where $\alpha,\beta,\gamma>0$ and $0\leq\theta<1$.\\
The survival function and hazard rate function of the EWG
distribution, are given respectively by
\begin{equation}\label{Survival}
S(y)=\frac{1-\left(1-e^{-(\beta
y)^{\gamma}}\right)^{\alpha}}{1-\theta\left(1-e^{-(\beta
y)^{\gamma}}\right)^{\alpha}},
\end{equation}
and
\begin{equation*}
h(y)=\frac{(1-\theta)\alpha\gamma\beta^{\gamma}y^{\gamma-1}e^{-(\beta y)^{\gamma}}
\left(1-e^{-(\beta y)^{\gamma}}\right)^{\alpha-1}}{\left[1-\theta\left(1-e^{-(\beta y)^
{\gamma}}\right)^{\alpha}\right]\left[1-\left(1-e^{-(\beta y)^{\gamma}}\right)^{\alpha}\right]}.
\end{equation*}

\begin{prop}
For $|\theta|<1$, it is easy to prove that the density of EWG
distribution can be written as an infinite mixture of EW
distributions. If $|z|< 1$ and $k > 0$, we have the series
representation
\begin{equation}\label{series}
(1-z)^{-k}=\sum_{j=0}^{\infty}\frac{\Gamma(k+j)}{\Gamma(k)j!}z^j.
\end{equation}
If $|\theta|<1$, expanding $\left[1-\theta\left(1-e^{-(\beta
y)^{\gamma}}\right)^{\alpha}\right]^{-2}$ as in Eq. (\ref{series}),
the density function (\ref{pdf EWG}) can be demonstrated by
\begin{equation*}
f_Y(y)=(1-\theta)\alpha\gamma\beta^{\gamma}y^{\gamma-1}e^{-(\beta
y)^{\gamma}}\sum_{j=0}^{\infty}(j+1)\theta^{j}\left(1-e^{-(\beta
y)^{\gamma}}\right)^{\alpha (j+1)-1}.
\end{equation*}
Using the EW density (\ref{pdf ExW}), we obtain
\begin{equation}\label{new EWG}
f_{EWG}(y;\alpha,\beta,\gamma,\theta)=(1-\theta)\sum_{j=0}^{\infty}\theta^{j}f_{EW}(y;\alpha(j+1),\beta,\gamma).
\end{equation}
Various mathematical properties of the EWG distribution for
$|\theta|<1$, can be obtained from Eq. (\ref{new EWG}) and the
corresponding properties of the EW distribution.
\end{prop}
\begin{prop}
The density of EWG distribution can be expressed as infinite linear
combination of density of the biggest order statistic of
$X_{1},\cdots,X_{n}$, where $X_{i}\sim EW(\alpha,\beta,\gamma)$ for
$i=1,2,\cdots,n$. we have
\begin{equation*}
f_{EWG}(y)=\sum^{\infty}_{n=1}(G(y))^{n}P(N=n)=\sum^{\infty}_{n=1}g_{X_{(n)}}(y)P(N=n),
\end{equation*}
in which $g_{X_{(n)}}(y)$ is the pdf of
$X_{(n)}=\max(X_{1},\cdots,X_{n})$.
\end{prop}

\section{ Quantiles and moments of the EWG distribution }
The $p$th quantile of the EWG distribution is given by
\begin{equation*}
x_{p}=\beta^{-1} \left\{ \left[-\log
\left(1-\left(\frac{p}{1-\theta(1-p)}\right)^{\frac{1}{\alpha}}\right)\right]^{\frac{1}{\gamma}}\right\},
\end{equation*}
which is used for data generation from the EWG distribution. In
particular, the median of the EWG distribution is given by
\begin{equation*}
x_{0.5}=\beta^{-1}\left\{\left[-\log
\left(1-\left(\frac{1}{2-\theta}\right)^{\frac{1}{\alpha}}\right)\right]^{\frac{1}{\gamma}}\right\}.
\end{equation*}

Suppose that $Y\sim EWG(\alpha,\beta,\gamma,\theta)$, and
$X_{(n)}=\max(X_{1},\cdots,X_{n})$, where $X_{i}\sim
EW(\alpha,\beta,\gamma)$ for $i=1,2,\cdots,n$,  then the $k$th
moment of $Y$ is given by
\begin{equation}\label{meank EWG}
\begin{array}[b]{ll}
E(Y^{k})=E(E(Y^{k}|N))&=\sum^{\infty}_{n=1}P(N=n)E(Y^{k}|n)=\sum^{\infty}_{n=1}P(N=n)E(X^{k}_{(n)})\medskip\\
&=\sum^{\infty}_{n=1}P(N=n)n\alpha \beta^{-k}\Gamma\left(\frac{k}{\gamma}+1\right)\sum^{\infty}_{j=0}(-1)^j {n\alpha-1 \choose j}(j+1)^
{-(\frac{k}{\gamma}+1)}\medskip\\
&=(1-\theta)\alpha
\beta^{-k}\Gamma\left(\frac{k}{\gamma}+1\right)\sum^{\infty}_{n=1}\sum^{\infty}_{j=0}n
\theta^{n-1}(-1)^j {n\alpha-1 \choose
j}(j+1)^{-(\frac{k}{\gamma}+1)}.
\end{array}
\end{equation}
For positive integer values of $\alpha$, the index $j$ in above
expression stops at $\alpha-1$. \newline Using Eq. (\ref{meank
EWG}), the moment generating function of the EWG distribution is
given by
\begin{equation}
\begin{array}[b]{l}\label{mgf EWP}
M_{Y}(t)=\sum^{\infty}_{i=0}\frac{t^{i}}{i!}E(Y^{i})\medskip\\
=\sum^{\infty}_{i=0}\frac{t^{i}}{i!}\left[(1-\theta)\alpha
\beta^{-i}\Gamma\left(\frac{i}{\gamma}+1\right)\sum^{\infty}_{n=1}\sum^{\infty}_{j=0}n
\theta^{n-1}(-1)^j {n\alpha-1 \choose j}(j+1)^{-(\frac{i}{\gamma}+1)}\right]\medskip\\
=\alpha(1-\theta)\sum^{\infty}_{i=0}\sum^{\infty}_{n=1}\sum^{\infty}_{j=0}n
\theta^{n-1}(-1)^j {n\alpha-1 \choose
j}\Gamma\left(\frac{i}{\gamma}+1\right)\frac{(t/\beta)^{i}}{i!}(j+1)^{-(\frac{i}{\gamma}+1)}.
\end{array}
\end{equation}
%Using $(\ref{a})$ for, $|t|<1$, the cumulant generating function of
%$Y$ is
%\begin{equation*}
%K_{Y}(t)=\log[M_{Y}(t)]=\log\Big\{(e^{\lambda}-1)^{-1}\sum^{\infty}_{n=1}\sum^{n}_{k=0}
%\frac{\lambda^{n}}{n!}\frac{t^{k}}{k!}(n\gamma)^{k/\beta}\alpha^{k}\Gamma(1-k\beta^{-1})\Big\}~~~|t|<
%1,k\leq\beta
%\end{equation*}
%The first four cumulants obtained as follow
%\begin{equation*}
%\begin{array}[b]{ll}
%k_{1}&={\lambda({e^{\lambda}-1})^{-1}\alpha\gamma^{\frac{1}{\beta}}}\Gamma(1-\beta^{-1})\\\\
%k_{2}&={\lambda({e^{\lambda}-1})^{-1}\alpha^{2}\gamma^{\frac{2}{\beta}}}\Big\{\Gamma(1-2\beta^{-1})
%-{\lambda}({e^{\lambda}-1})^{-1}\Gamma^{2}(1-\beta^{-1})\Big\}\\\\
%k_{3}&={\lambda({e^{\lambda}-1})^{-1}\alpha^{3}\gamma^{\frac{3}{\beta}}}\Big\{\Gamma(1-3\beta^{-1})-{3\lambda}
%({e^{\lambda}-1})^{-1}\Gamma(1-2\beta^{-1})\Gamma(1-\beta^{-1})+{2\lambda^{2}}{(e^{\lambda}-1)^{-2}}\Gamma^{3}(1-\beta^{-1})\Big\}
%\end{array}
%\end{equation*}
According to Eq. (\ref{meank EWG}), the mean and variance of the EWG
distribution are given respectively by
\begin{equation}\label{mean EWG}
E(Y)=(1-\theta)\alpha
\beta^{-1}\Gamma\left(\frac{1}{\gamma}+1\right)
\sum^{\infty}_{n=1}\sum^{\infty}_{j=0}n \theta^{n-1}(-1)^j
{n\alpha-1 \choose j}(j+1)^{-(\frac{1}{\gamma}+1)},
\end{equation}
and
\begin{equation}\label{var EWG}
Var(Y)=(1-\theta)\alpha
\beta^{-2}\Gamma\left(\frac{2}{\gamma}+1\right)\sum^{\infty}_{n=1}\sum^{\infty}_{j=0}n
\theta^{n-1}(-1)^j {n\alpha-1 \choose
j}(j+1)^{-(\frac{2}{\gamma}+1)}-E^{2}(Y),
\end{equation}
Where $E(Y)$ is given in Eq. (\ref{mean EWG})
%\begin{prop} The mgf
%and the $k$th moment of the EWG distribution can be obtained using
%the following expressions:
%\begin{equation*}
%M_{Y}(t)=E\left[\sum^{N}_{k=0}\frac{t^{k}}{k!}(N\gamma)^{k/\beta}\alpha^{k}
%\Gamma(1-k\beta^{-1})\right],~~~~k< \beta,~|t|<1,
%\end{equation*}
%and
%\begin{equation*}
%\mu_{k}=E(Y^{k})=\alpha^{k}\gamma^{k/\beta}\Gamma(1-k\beta^{-1})E[N^{k/\beta}],~~~~k<
%\beta,
%\end{equation*}
%%\begin{equation*}
%E[N^{k/\beta}]=\lambda(e^{\lambda}-1)^{-1}\sum^{\infty}_{j=0}\frac{\lambda^{j}(j+1)^{\frac{k}{\beta}-1}}{j!}.
%\end{equation*}
%\end{prop}
\section{ R\'{e}nyi and Shannon entropies}
For a random variable with pdf $f$, the R\'{e}nyi entropy is defined
by $I_{R}(r)=\frac{1}{1-r}\log\{\int_{\mathbb{R}}f^{r}(y)dy\}$, for
$r>0$ and $r\neq 1$. For the EWG distribution, the power series
expansion gives
\begin{equation*}
\begin{array}[b]{ll}
\int^{\infty}_{0} f^{r}(y)dy&=\left[\alpha\gamma\beta^{\gamma}
(1-\theta)\right]^{r}\sum_{j=0}^{\infty}\theta^{j}\frac{\Gamma(2r+j)}{\Gamma(2r)j!}\int^{\infty}_{0}y^{(\gamma-1)r}e^{-r(\beta
y)^\gamma}
(1-e^{-(\beta y)^\gamma})^{\alpha( j+r)-r}\medskip\\
&=\left[\alpha\gamma\beta^{\gamma}(1-\theta)\right]^{r}\sum_{j=0}^{\infty}\sum_{k=0}^{\infty}(-1)^k\theta^{j}\frac{\Gamma(2r+j)}{\Gamma(2r)j!}
{\alpha (j+r)-r \choose k}\int^{\infty}_{0}
y^{(\gamma-1)r}e^{-(k+r)(\beta y)^\gamma}.
\end{array}
\end{equation*}
Setting $u=(\beta y)^\gamma$, gives
\begin{equation}\label{f^r}
\begin{array}[b]{ll}
\int^{\infty}_{0} f^{r}(y)dy&=\left[\alpha
(1-\theta)\right]^{r}\left(\beta\gamma\right)^{r-1}
\sum_{j=0}^{\infty}\sum_{k=0}^{\infty}(-1)^k{\alpha (j+r)-r \choose
k}\frac{\Gamma(2r+j)}{\Gamma(2r)j!}
\frac{\Gamma(r-\frac{r-1}{\gamma})}{(k+r)^{r-\frac{r-1}{\gamma}}}\theta^{j}
\end{array}.
\end{equation}
Substituting from (\ref{f^r}), the R\'{e}nyi entropy is given by
\begin{equation}
I_{R}(r)=\frac{1}{1-r} \log \left\{\left[\alpha
(1-\theta)\right]^{r}\left(\beta\gamma\right)^{r-1}
\sum_{j=0}^{\infty}\sum_{k=0}^{\infty}(-1)^k{\alpha (j+r)-r \choose
k}\frac{\Gamma(2r+j)}{\Gamma(2r)j!}
\frac{\Gamma(r-\frac{r-1}{\gamma})}{(k+r)^{r-\frac{r-1}{\gamma}}}\theta^{j}\right\}.
\end{equation}
The Shannon entropy which is defined by $E[-\log (f(Y))]$, is
derived from  $\lim_{r\rightarrow 1}I_{R}(r)$.

\section{ Moments of order statistics}
Let the random variable $Y_{r:n}$ denotes the $r$th order statistic
$(Y_{1:n}\leq Y_{2:n}\leq \cdots\leq Y_{n:n})$ in a sample of size
$n$ from the EWG  distribution. The pdf of $Y_{r:n}$ for
$r=1,\cdots,n$, is given by
\begin{equation}\label{ordr EWG}
f_{r:n}(y)=\frac{1}{B(r,n-r+1)}f(y)F(y)^{r-1}[1-F(y)]^{n-r},~~~y>0.
\end{equation}
where $F(y)$ and f(y) are the cdf and pdf of the random variable
$Y$. Substituting from (\ref{cdf EWG}) and (\ref{pdf EWG}) into
(\ref{ordr EWG}), gives

\begin{equation}
\begin{array}[b]{ll}\label{f_{r:n}}
f_{r:n}(y)&=\frac{\alpha \gamma
\beta^{\gamma}(1-\theta)^{r}}{B(r,n-r+1)}y^{\gamma-1}e^{-(\beta
y)^\gamma} \frac{(1-e^{-(\beta y)^\gamma})^{\alpha
r-1}\left[1-\left(1-e^{-(\beta y)^{\gamma}}\right)^{\alpha}
\right]^{n-r}} {\left[1-\theta \left(1-e^{-(\beta
y)^{\gamma}}\right)^{\alpha} \right]^{n+1}},~~~y>0.
\end{array}
\end{equation}
 Also the cdf of $Y_{r:n}$ is
 given by
\begin{equation}\label{F_{r:n}}
\begin{array}[b]{ll}
F_{r:n}(y)&=\sum^{n}_{k=r}{{n}\choose{k}}[F(y)]^{k}[1-F(y)]^{n-k}\medskip\\&=\sum^{n}_{k=r}{{n}\choose{k}}\frac{(1-\theta)^{k}\left(1-e^{-(\beta
y)^{\gamma}}\right)^{\alpha k}\left(1-\left(1-e^{-(\beta
y)^{\gamma}}\right)^{\alpha}\right)^{n-k}}{\left(1-\theta\left(1-e^{-(\beta
y)^{\gamma}}\right)^{\alpha}\right)^{n}}.
\end{array}
\end{equation}

Using the binomial expansion, series expansion (\ref{series}) and
after some calculations, the $k$th moment of the $r$th order
statistic $Y_{r:n}$ is given by
\begin{equation}\label{E(Y_{r:n}}
E(Y_{r:n} ^{k})=\frac{\alpha
\beta^{-k}(1-\theta)^{r}}{B(r,n-r+1)}\sum^{\infty}_{i=0}\sum^{n-r}_{j=0}
\sum^{\infty}_{s=0}(-1)^{j+s}\theta^{i}{{n+i}\choose{i}}{{n-r}\choose{j}}{{\alpha(i+j+r)-1}\choose{s}}
\frac{\Gamma(\frac{k}{\gamma}+1)}{(s+1)^{\frac{k}{\gamma}+1}}.
\end{equation}
The pdf, cdf and $k$th moment of the smallest and biggest order
statistics, i.e., $Y_{1:n}$ and $Y_{n:n}$, can be obtained by
setting $r=1$ and $n$ in Eqs. (\ref{f_{r:n}})-(\ref{E(Y_{r:n}}).

\section{ Residual life function of the EWG distribution}
Given that a component survives up to time $t\geq0$, the residual
life is the period beyond $t$ until the time of failure and defined
by expectation of the conditional random variable $X|X > t$. In
reliability, it is well known that the mean residual life function
and ratio of two consecutive moments of residual life, determine the
distribution uniquely (Gupta and Gupta, \cite{Gupta1983}).
Therefore, we obtain the $r$th order moment of the residual life via
the general formula
\begin{equation}\label{m_{r}(t)}
m_{r}(t)=E\left[(Y-t)^r|Y>t\right]=\frac{1}{S(t)}\int_{t}^{\infty}(y-t)^{r}f(y)dy,
\end{equation}
where $S(t)=1-F(t)$, is the survival function.
\newline In what seen this onwards, we use the expressions
\begin{equation*}\label{incom int1}
\int_{t}^{\infty}x^{\gamma+s-1}e^{-(k+1)(\beta
x)^{\gamma}}dx=\frac{1}{\gamma\beta^{\gamma+s}}(k+1)^{-(1+\frac{s}{\gamma})}\Gamma^{(k+1)(\beta
t)^{\gamma}}(1+\frac{s}{\gamma}),
\end{equation*}
and
\begin{equation*}\label{incom int2}
\int_{0}^{t}x^{\gamma+s-1}e^{-(k+1)(\beta
x)^{\gamma}}dx=\frac{1}{\gamma\beta^{\gamma+s}}(k+1)^{-(1+\frac{s}{\gamma})}\Gamma_{(k+1)(\beta
t)^{\gamma}}(1+\frac{s}{\gamma}),
\end{equation*}
where $\Gamma^{t}(s)= \int_{t}^{\infty}x^{s-1}e^{-x}dx$ is the upper
incomplete gamma function and $\Gamma_{t}(s)=
\int_{0}^{t}x^{s-1}e^{-x}dx$ is the lower incomplete gamma function.
\newline Applying series expansion (\ref{series}), the binomial expansion for
$(y-t)^r$ and substituting $S(y)$ given by (\ref{Survival}) into
(\ref{m_{r}(t)}), the $r$th moment of the residual life of the EWG
is given by
\begin{equation}\label{r res}
%\begin{array}[b]{ll}
m_{r}(t)=\frac{\alpha(1-\theta)}{S(t)}\sum_{i=0}^{r}\sum_{j=0}^{\infty}\sum_{k=0}^{\infty}\frac{(-1)^{i+k}(j+1)t^{i}\theta^{j}}{(k+1)^
{1+\frac{r-i}{\gamma}}\beta^{r-i}}
{{r}\choose{i}}{{\alpha(j+1)-1}\choose{k}}\Gamma^{(k+1)(\beta
t)^{\gamma}}\left(1+\frac{r-i}{\gamma}\right).
%\end{array}
\end{equation}
Another important representation for the EWG is the mean Residual
life (MRL) function obtain by setting $r=1$ in Eq. (\ref{r res}).
The importance of the MRL function is due to its uniquely
determination of the lifetime distribution as well as the failure
rate (FR) function. Lifetimes can exhibit IMRL (increasing MRL) or
DMRL (decreasing MRL). MRL functions that first decreases
(increases) and then increases (decreases) are usually called
bathtub (upside-down bathtub) shaped, BMRL (UMRL). Many authors such
as Ghitany \cite{Ghitany}, Mi \cite{Mi}, Park \cite{Park } and Tang
et al. \cite{Tang } have been studied the relationship between the
behaviors of the MLR and FR functions of a distribution. \newline
The following theorem gives the MRL function of the EWG
distribution.
\begin{thm}
The MRL function of the EWG distribution is given by
\begin{equation*}\label{MRL EWG}
m_{1}(t)=\Big[\frac{\alpha(1-\theta)}{\beta
S(t)}\sum_{i=0}^{\infty}\sum_{j=0}^{\infty}(-1)^{j}{{\alpha(i+1)-1}\choose{j}}\theta^{i}(i+1)(j+1)^{-(1+\frac{1}{\gamma})}
\Gamma^{(j+1)(\beta t)^{\gamma}}(1+\frac{1}{\gamma})\Big]-t.
\end{equation*}
\end{thm}
Setting $r=2$ in (\ref{r res}), the variance of residual life
function of the EWG distribution can be obtained using $m_{1}(t)$
and $m_{2}(t)$.

%\begin{thm}
%The MRL function of the EWG distribution with cdf (\ref{cdf EWG}) is
%\begin{equation}\label{MRL EWG}
%m(t)=(\mu_{1}+I(t)-t)/S(t),~~t\geq 0
%\end{equation}
%where $I(t)=\int^{t}_{0}F(y)dy$, $S(t)$ is the survival function
%given in (), and $\mu_{1}$ is the mean of the EWG in Eq. (\ref{mean
%EWG}).
%\end{thm}
%\begin{pf}
%For more detail about the proof of this theorem see Nassar and Eissa
%(2003).
%\end{pf}
%According to Theorem 1, for the EWG distribution with $f(y)$ given
%by (\ref{pdf EWG}), we have
%\begin{equation}\label{int MRL EWG}
%I(t)=\frac{1-\theta}{\beta\gamma}\sum_{j=0}^{\infty}\sum_{l=0}^{\infty}\frac{(-1)^{l+j}\theta^{j}}{l^{1/\gamma}}{{\alpha(j+1)}\choose{l}}
%\Psi(1/\gamma;l(\beta t)^{\gamma}),
%\end{equation}
%where $\Psi(s; t)$ is the lower incomplete gamma function given by
%$\Psi(s; t)= \int_{0}^{t}x^{s-1}e^{-x}dx$. Substituting Eqs.
%(\ref{mean EWG}), () and (\ref{int MRL EWG}) into (\ref{MRL EWG})
%gives the MRL function of the EWG distribution.

%\begin{equation*}
%\begin{array}[b]{ll}
%m(t) &= \alpha (1-\theta)\sum^{\infty}_{n=1}\sum^{n\alpha-1}_{j=0}\frac{n\theta^{n-1}(-1)^{j}{n\alpha-1 \choose j}}{(j+1)S(t)}\{\beta^{-1}[ (j+1)^{-\frac{1}{\gamma}}\Gamma\left(\frac{1}{\gamma}+1\right)\medskip \\
%&- \frac{(j+1)^{-\frac{1}{\gamma}}}{\gamma} \Gamma_{(j+1)\tau} \left(\frac{1}{\gamma}\right) ] +t e^{-(j+1)\tau}\}-t
%\end{array}
%\end{equation*}
%Where $\tau=(t\beta)^{\gamma}$ and $\Gamma_{a}(r)=\int^{0}_{\infty}u^{r-1}e^{-u}du $ is the incomplets gamma function.

\section{ Estimation and inference}
In this section, we study the estimation of the parameters of the
EWG distribution. Let $Y_{1},Y_{2},\cdots,Y_{n}$ be a random sample
with observed values $y_{1},y_{2},\cdots,y_{n}$ from EWG
distribution with parameters $\alpha,\beta,\gamma$ and $\theta$. Let
$\Theta=(\alpha,\beta,\gamma,\theta)^{T}$ be the parameter vector.
The total log-likelihood function is given by
\begin{equation*}
\begin{array}[b]{ll}
l_{n}\equiv l_{n}(y;\Theta)&= n[\log \alpha +\log \gamma + \gamma \log \beta + \log (1-\theta)]+ (\gamma-1)\sum^{n}_{i=1}
\log y_{i}-\sum^{n}_{i=1}(\beta y_{i})^{\gamma}\medskip \\
&~~~ +(\alpha-1)\sum^{n}_{i=1}\log (1-e^{-(\beta
y_{i})^{\gamma}})-2\sum^{n}_{i=1}\log[1-\theta\left(1-e^{-(\beta
y_{i})^{\gamma}}\right)^{\alpha}].
\end{array}
\end{equation*}
The associated score function is given by $U_{n}(\Theta)=(\partial
l_{n}/\partial \alpha,\partial l_{n}/\partial \beta, \partial
l_{n}/\partial \gamma,\partial l_{n}/\partial \lambda)^{T}$, where

\begin{equation*}
\begin{array}{ll}
\frac{\partial l_{n}}{\partial
\alpha}&=\frac{n}{\alpha}+\sum^{n}_{i=1}\log (1-e^{-(\beta y_{i})^{\gamma}})+2\theta \sum^{n}_{i=1}\log (1-e^{-(\beta y_{i})
^{\gamma}})\frac{(1-e^{-(\beta y_{i})^{\gamma}})^{\alpha}}{1-\theta(1-e^{-(\beta y_{i})^{\gamma}})^{\alpha}},\medskip \\

\frac{\partial l_{n}}{\partial \beta}&=\frac{n\gamma}{\beta}-\gamma\beta^{\gamma-1} \sum^{n}_{i=1}y_{i}^{\gamma}+
(\alpha-1)\gamma\beta^{\gamma-1} \sum^{n}_{i=1}\frac{y_{i}^{\gamma}e^{-(\beta y_{i})^{\gamma}}}{1-e^{-(\beta y_{i})^{\gamma}}}\medskip \\
&~~+2\theta\alpha\gamma\beta^{\gamma-1}\sum^{n}_{i=1}\frac{y_{i}^{\gamma}e^{-(\beta
y_{i})^{\gamma}}(1-e^{-(\beta y_{i})^{\gamma}})
^{\alpha-1}}{1-\theta(1-e^{-(\beta y_{i})^{\gamma}})^{\alpha}},\medskip \\

\frac{\partial l_{n}}{\partial
\gamma}&=\frac{n}{\gamma}+n\log\beta+\sum^{n}_{i=1}\log
y_{i}-\sum^{n}_{i=1}\log(\beta y_{i})(\beta y_{i})^{\gamma}
\medskip \\
 &~~+(\alpha-1)\sum^{n}_{i=1}\frac{\log(\beta y_{i})(\beta y_{i})^{\gamma}e^{-(\beta y_{i})^{\gamma}}}{1-e^{-(\beta y_{i})^{\gamma}}}
 + 2\theta \alpha \sum^{n}_{i=1}\frac{\log(\beta y_{i})(\beta y_{i})^{\gamma}e^{-(\beta y_{i})^{\gamma}}(1-e^{-(\beta y_{i})
 ^{\gamma}})^{\alpha-1}}{1-\theta(1-e^{-(\beta y_{i})^{\gamma}})^{\alpha}} ,\medskip \\

\frac{\partial l_{n}}{\partial
\theta}&=-\frac{n}{1-\theta}+2\sum^{n}_{i=1}\frac{(1-e^{-(\beta
y_{i})^{\gamma}})^{\alpha}}{1-\theta(1-e^{-(\beta
y_{i})^{\gamma}})^{\alpha}}.
\end{array}
\end{equation*}
The maximum likelihood estimation (MLE) of $\Theta $, say
$\widehat{\Theta }$, is obtained by solving the nonlinear system
$U_n\left(\Theta \right)=\textbf{0}$. The solution of this nonlinear
system of equation has not a closed form. For interval estimation
and hypothesis tests on the model parameters, we require the
information matrix. The $4\times 4$ observed information matrix is
\[I_n\left(\Theta \right)=-\left[ \begin{array}{cccc}
I_{\alpha \alpha } & I_{\alpha \beta } & I_{\alpha \gamma } & I_{\alpha \theta }\\
I_{\alpha \beta } & I_{\beta \beta } & I_{\beta \gamma }& I_{\beta \theta } \\
I_{\alpha \gamma } & I_{\beta \gamma } & I_{\gamma \gamma}& I_{\gamma \theta } \\
I_{\alpha \theta } & I_{\beta \theta } & I_{\gamma \theta}& I_{\theta \theta } \\
\end{array} \right],\] whose elements are given in Appendix.

Applying the usual large sample approximation, MLE of $\Theta $ i.e.
$\widehat{\Theta }$ can be treated as being approximately
$N_4(\Theta ,{J_n(\Theta )}^{-1}{\mathbf )}$, where $J_n\left(\Theta
\right)=E\left[I_n\left(\Theta \right)\right]$. Under conditions
that are fulfilled for parameters in the interior of the parameter
space but not on the boundary, the asymptotic distribution of
$\sqrt{n}(\widehat{\Theta }{\rm -}\Theta {\rm )}$ is $N_4({\mathbf
0},{J(\Theta )}^{-1})$, where $J\left(\Theta \right)={\mathop{\lim
}_{n\to \infty } {n^{-1}I}_n(\Theta )\ }$ is the unit information
matrix. This asymptotic behavior remains valid if $J(\Theta )$ is
replaced by the average sample information matrix evaluated at
$\widehat{\Theta }$, say ${n^{-1}I}_n(\widehat{\Theta })$. The
estimated asymptotic multivariate normal $N_4(\Theta
,{I_n(\widehat{\Theta })}^{-1})$ distribution of $\widehat{\Theta }$
can be used to construct approximate confidence intervals for the
parameters and for the hazard rate and survival functions. An
$100(1-\gamma )$ asymptotic confidence interval for each parameter
${\Theta }_{{\rm r}}$ is given by
\[{ACI}_r=({\widehat{\Theta
}}_r-Z_{\frac{\gamma }{2}}\sqrt{{\hat{I}}^{rr}},{\widehat{\Theta
}}_r+Z_{\frac{\gamma }{2}}\sqrt{{\hat{I}}^{rr}}),\] where
${\hat{I}}^{rr}$ is the $(\textit{r, r})$ diagonal element of
${I_{n}(\widehat{\Theta })}^{-1}$ for $r=1,~2,~3,~4,$ and
$Z_{\frac{\gamma }{2}}$ is the quantile $1-\gamma /2$ of the
standard normal distribution.

%\begin{figure}[t]
%\centering
%\includegraphics[scale=0.33]{g11.eps}
%\includegraphics[scale=0.33]{g22.eps}
%\includegraphics[scale=0.33]{g33.eps}
%\includegraphics[scale=0.33]{g44.eps}
%\includegraphics[scale=0.33]{g55.eps}
%\includegraphics[scale=0.33]{g66.eps}
%\includegraphics[scale=0.33]{g77.eps}
%\includegraphics[scale=0.33]{g88.eps}
%\includegraphics[scale=0.33]{g99.eps}
%\caption[]{Plots of density and hazard rate functions of GIW
%distribution for different values $\alpha $, $\beta $, and $\theta
%$.}
%\end{figure}

%\begin{figure}[t]
%\centering
%\includegraphics[scale=0.33]{GIWP(pdf1).eps}
%\includegraphics[scale=0.33]{GIWP(pdf2).eps}
%\includegraphics[scale=0.33]{GIWP(pdf3).eps}
%\includegraphics[scale=0.33]{GIWP(pdf7).eps}
%\includegraphics[scale=0.33]{GIWP(pdf8).eps}
%\includegraphics[scale=0.33]{GIWP(pdf9).eps}
%\caption[]{Plots of density function of GIWP distribution for
%different values $(\alpha,~\beta,~\theta,~\lambda) $.}
%\end{figure}

%\begin{figure}[t]
%\centering
%\includegraphics[scale=0.33]{GIWP(hazard1).eps}
%\includegraphics[scale=0.33]{GIWP(hazard1).eps}
%\includegraphics[scale=0.33]{GIWP(hazard3).eps}
%\includegraphics[scale=0.33]{GIWP(hazard7).eps}
%\includegraphics[scale=0.33]{GIWP(hazard8).eps}
%\includegraphics[scale=0.33]{GIWP(hazard9).eps}
%\caption[]{Plots of hazard rate function of GIWP distribution for
%different values $(\alpha,~\beta,~\theta,~\lambda) $.}
%\end{figure}

\section{Submodels of the EWG distribution   }
The EWG distribution contains some sub-models for the special values
of $\alpha$, $\beta$ and $\gamma$. Some of these distributions are
discussed here in details.

\subsection{Complementary Weibull-geometric distribution}
The complementary Weibull-geometric (CWG) distribution is a special
case of the EWG distribution for $\alpha=1$. Our approach here is
complementary to that of Barreto-Souza et al.
\cite{Barreto-Souza2011 } in introducing the Weibull-geometric (WG)
distribution. The pdf, cdf and hazard rate function of the CWG
distribution, are given respectively by
\begin{equation}\label{pdf WG}
f(y)=\frac{(1-\theta)\gamma\beta^{\gamma}y^{\gamma-1}e^{-(\beta y)^{\gamma}}}{\left[1-\theta\left(1-e^{-(\beta y)^{\gamma}}\right)\right]^2},
\end{equation}
\begin{equation}\label{cdf WG}
F(y)=\frac{(1-\theta)\left(1-e^{-(\beta
y)^{\gamma}}\right)}{1-\theta\left(1-e^{-(\beta
y)^{\gamma}}\right)},
\end{equation}
and
\begin{equation}\label{hazard WG}
h(y)=\frac{(1-\theta)\gamma\beta^{\gamma}y^{\gamma-1}e^{-(\beta y)^{\gamma}}}{\left[1-\theta\left(1-e^{-(\beta y)^{\gamma}}\right)\right]\left[1-\left(1-e^{-(\beta y)^{\gamma}}\right)\right]}.
\end{equation}
According to Eqs. (\ref{mean EWG}) and (\ref{var EWG}), the mean and
variance of the CWG distribution are given by
\begin{equation}\label{ExCWG}
E(Y)=\frac{(1-\theta)}{\beta}
\Gamma\left(1+\frac{1}{\gamma}\right)\sum^{\infty}_{n=1}\sum^{\infty}_{j=0}n
\theta^{n-1}(-1)^j {n-1 \choose j}(j+1)^{-(1+\frac{1}{\gamma})},
\end{equation}
and
\begin{equation}\label{VxCWG}
Var(Y)=\frac{(1-\theta)}{\beta^2}\Gamma\left(1+\frac{2}{\gamma}\right)\sum^{\infty}_{n=1}\sum^{\infty}_{j=0}n
\theta^{n-1}(-1)^j {n-1 \choose j}(j+1)^{-(1+\frac{2}{\gamma})}
-E^{2}(Y),
\end{equation}
where $E(Y)$ is given in Eq. (\ref{ExCWG}).

\subsection{Generalized exponential-geometric distribution  }

The generalized exponential-geometric (GEG) distribution is a
special case of the EWG distribution for $\gamma$=1. This
distribution is introduced and analyzed in details by Mahmoudi and
Jafari \cite{Mahmoudi2011a }. The pdf, cdf and hazard rate function
of the GEG distribution, are given respectively by
\begin{equation*}\label{pdf GEG}
f(y)=\frac{(1-\theta)\alpha\beta e^{-(\beta y)}\left(1-e^{-(\beta
y)}\right)^{\alpha-1}}{\left[1-\theta\left(1-e^{-(\beta
y)}\right)^{\alpha}\right]^2},
\end{equation*}
\begin{equation*}\label{cdf GEG}
F(y)=\frac{(1-\theta)\left(1-e^{-(\beta
y)}\right)^{\alpha}}{1-\theta\left(1-e^{-(\beta
y)}\right)^{\alpha}},
\end{equation*}
and
\begin{equation*}\label{hazard GEG}
h(y)=\frac{(1-\theta)\alpha\beta e^{-(\beta y)}\left(1-e^{-(\beta
y)}\right)^{\alpha-1}}{\left[1-\theta\left(1-e^{-(\beta
y)}\right)^{\alpha}\right]\left[1-\left(1-e^{-(\beta
y)}\right)^{\alpha}\right]}.
\end{equation*}
According to Eqs. (\ref{mean EWG}) and (\ref{var EWG}), the mean and
variance of the GEG distribution are
\begin{equation}\label{ExGEG}
E(Y)=\frac{\alpha(1-\theta)}
{\beta}\sum^{\infty}_{n=1}\sum^{\infty}_{j=0}n \theta^{n-1}(-1)^j
{n\alpha-1 \choose j}(j+1)^{-2},
\end{equation}
and
\begin{equation*}
Var(Y)=\frac{2\alpha(1-\theta)}
{\beta^2}\sum^{\infty}_{n=1}\sum^{\infty}_{j=0}n \theta^{n-1}(-1)^j
{n\alpha-1 \choose j}(j+1)^{-3}-E^{2}(Y),
\end{equation*}
where $E(Y)$ is given in Eq. (\ref{ExGEG}).

\subsection{Complementary exponential-geometric distribution}
The complementary exponential-geometric (CEG) distribution is a
special case of the EWG distribution for $\alpha=1$ and $\gamma=1$.
The pdf, cdf and hazard rate function of the CEG distribution are
given respectively by
\begin{equation*}\label{pdf EG}
f(y)=\frac{(1-\theta)\beta e^{-(\beta y)}}{\left[1-\theta\left(1-e^{-(\beta y)}\right)\right]^2},
\end{equation*}
\begin{equation*}\label{cdf EG}
F(y)=\frac{(1-\theta)\left(1-e^{-(\beta y)}\right)}{1-\theta\left(1-e^{-(\beta y)}\right)},
\end{equation*}
and
\begin{equation*}\label{hazard EG}
h(y)=\frac{(1-\theta)\beta e^{-(\beta y)}}{\left[1-\theta\left(1-e^{-(\beta y)}\right)\right]\left[1-\left(1-e^{-(\beta y)}\right)\right]}.
\end{equation*}
According to Eqs. (\ref{mean EWG}) and (\ref{var EWG}), the mean and
variance of the CEG distribution, are given respectively by
\begin{equation}\label{ExCEG}
E(Y)=(1-\theta) \beta^{-1}\sum^{\infty}_{n=1}\sum^{\infty}_{j=0}n
\theta^{n-1}(-1)^j {n-1 \choose j}/(j+1)^{2},
\end{equation}
and
\begin{equation*}
Var(Y)=(1-\theta)
\beta^{-2}\Gamma(3)\sum^{\infty}_{n=1}\sum^{\infty}_{j=0}n
\theta^{n-1}(-1)^j {n-1 \choose j}/(j+1)^{3}-E^{2}(Y),
\end{equation*}
where $E(Y)$ is given in Eq. (\ref{ExCEG}).

\subsection{Exponentiated Rayleigh-geometric distribution}
The exponentiated Rayleigh-geometric distribution (ERG) distribution
is a special case of the EWG distribution for $\gamma=2$. The pdf,
cdf and hazard rate function of the ERG distribution are given
respectively by
\begin{equation*}\label{pdf EG}
f_Y(y)=\frac{2(1-\theta)\alpha\beta^{2}y e^{-(\beta
y)^{2}}\left(1-e^{-(\beta
y)^{2}}\right)^{\alpha-1}}{\left[1-\theta\left(1-e^{-(\beta
y)^{2}}\right)^{\alpha}\right]^2},
\end{equation*}
\begin{equation*}\label{cdf EG}
F_Y(y)=\frac{(1-\theta)\left(1-e^{-(\beta
y)^{2}}\right)^{\alpha}}{1-\theta\left(1-e^{-(\beta
y)^{2}}\right)^{\alpha}},
\end{equation*}
and
\begin{equation*}\label{hazard EG}
h(y)=\frac{2(1-\theta)\alpha \beta^{2}y e^{-(\beta y)^{2}}
\left(1-e^{-(\beta y)^{2}}\right)^{\alpha-1}}{\left[1-\theta\left(1-e^{-(\beta y)^
{2}}\right)^{\alpha}\right]\left[1-\left(1-e^{-(\beta y)^{2}}\right)^{\alpha}\right]}.
\end{equation*}
The mean and variance of the ERG distribution, are given
respectively by
\begin{equation}\label{ExERG}
E(Y)=(1-\theta)\alpha
\beta^{-1}\Gamma\left(\frac{1}{2}+1\right)\sum^{\infty}_{n=1}\sum^{\infty}_{j=0}n
\theta^{n-1}(-1)^j {n\alpha-1 \choose j}(j+1)^{-(\frac{1}{2}+1)},
\end{equation}
and
\begin{equation*}
Var(Y)=(1-\theta)\alpha
\beta^{-2}\sum^{\infty}_{n=1}\sum^{\infty}_{j=0}n \theta^{n-1}(-1)^j
{n\alpha-1 \choose j}(j+1)^{-2}-E^{2}(Y),
\end{equation*}
where $E(Y)$ is given in Eq. (\ref{ExERG}).

\subsection{Rayleigh-geometric distribution}
The Rayleigh-geometric distribution (RG) distribution is a special
case of the CWG distribution, obtained by choosing $\gamma=2$ in CWG
distribution. Setting $\gamma=2$ in Eqs. (\ref{pdf WG})-(\ref{hazard
WG}) gives the pdf, cdf and hazard rate function of the RG
distribution as
\begin{equation*}\label{pdf RG}
f(y)=\frac{2(1-\theta)\beta^{2}y e^{-(\beta
y)^{2}}}{\left[1-\theta\left(1-e^{-(\beta
y)^{2}}\right)\right]^2} ,
\end{equation*}
\begin{equation*}\label{cdf RG}
F(y)= \frac{(1-\theta)\left(1-e^{-(\beta
y)^{2}}\right)}{1-\theta\left(1-e^{-(\beta
y)^{2}}\right)},
\end{equation*}
and
\begin{equation*}\label{hazard RG}
h(y)=\frac{2(1-\theta)\beta^{2}y e^{-(\beta
y)^{2}}}{e^{-(\beta
y)^{2}}\left[1-\theta\left(1-e^{-(\beta
y)^{2}}\right)\right]},
\end{equation*}
According to Eqs. (\ref{ExCWG}) and (\ref{VxCWG}), the mean and
variance of the RG distribution are given by
\begin{equation}\label{ExRG}
E(Y)=(1-\theta)
\beta^{-1}\Gamma\left(\frac{1}{2}+1\right)\sum^{\infty}_{n=1}\sum^{\infty}_{j=0}n
\theta^{n-1}(-1)^j {n-1 \choose j}/(j+1)^{-(\frac{1}{2}+1)},
\end{equation}
and
\begin{equation*}
Var(Y)=(1-\theta) \beta^{-2}\sum^{\infty}_{n=1}\sum^{\infty}_{j=0}n
\theta^{n-1}(-1)^j {n-1 \choose j}/(j+1)^{-2}-E^{2}(Y),
\end{equation*}
where $E(Y)$ is given in Eq. (\ref{ExRG}).

\section{ Conclusion}
We propose a new four-parameter distribution, referred to as the EWG
distribution which contains as special sub-models the generalized
exponential-geometric (GEG), complementary Weibull-geometric (CWG),
complementary exponential-geometric (CEG), exponentiated
Rayleigh-geometric (ERG) and Rayleigh-geometric (RG) distributions.
The hazard function of the EWG distribution can be decreasing,
increasing, bathtub-shaped and unimodal. Several properties of the
EWG distribution such as quantiles and moments, maximum likelihood
estimation procedure via an EM-algorithm, R\'{e}nyi and Shannon
entropies, moments of order statistics, residual life function and
probability weighted moments are studied. Finally, we fitted EWG
model to two real data sets to show the potential of the new
proposed distribution.

\end{document}